# The inconsistency of the *h*-index

Ludo Waltman and Nees Jan van Eck

Centre for Science and Technology Studies, Leiden University, The Netherlands
{waltmanlr, ecknjpvan}@cwts.leidenuniv.nl

The *h*-index is a popular bibliometric indicator for assessing individual scientists. We criticize the *h*-index from a theoretical point of view. We argue that for the purpose of measuring the overall scientific impact of a scientist (or some other unit of analysis) the *h*-index behaves in a counterintuitive way. In certain cases, the mechanism used by the *h*-index to aggregate publication and citation statistics into a single number leads to inconsistencies in the way in which scientists are ranked. Our conclusion is that the *h*-index cannot be considered an appropriate indicator of a scientist's overall scientific impact. Based on recent theoretical insights, we discuss what kind of indicators can be used as an alternative to the *h*-index. We pay special attention to the highly cited publications indicator. This indicator has a lot in common with the *h*-index, but unlike the *h*-index it does not produce inconsistent rankings.

## 1. Introduction

The introduction of the *h*-index (or Hirsch index) in 2005 has had an enormous influence on bibliometric and scientometric research. As can be seen in Figure 1, in 2010 and 2011 almost one out of four publications in *Scientometrics* and *Journal of Informetrics* cited the paper in which physicist Jorge E. Hirsch proposed his index (Hirsch, 2005). A large part of the literature building on Hirsch' work is concerned with introducing variants, extensions, and generalizations of the *h*-index. In a recent study (Bornmann, Mutz, Hug, & Daniel, 2011), no less than 37 variants of the *h*-index were listed.

Research into the development of new *h*-index variants and, more generally, of new bibliometric indicators often proceeds in a somewhat ad hoc fashion (see also Marchant, 2009b). Researchers take an indicator, identify a property of the indicator which they argue is undesirable, and then propose a new indicator which does not have this undesirable property. The weakness of this approach to indicator development is that in most cases it is unsystematic. The choice of a new indicator is often made in a somewhat arbitrary way, and there usually is no clear overall picture of the properties of the new indicator and of the way in which the indicator compares with existing indicators.

The literature on bibliometric indicators, and in particular on the *h*-index and its variants, is also strongly empirically oriented. For instance, new indicators are often justified mainly based on empirical grounds, by arguing that the results produced by an indicator are in agreement with what appears to be intuitively reasonable. Similarly, comparisons between indicators are often performed empirically, for instance by analyzing the strength of the correlation between indicators (e.g., Bornmann et al., 2011). Instead of studying indicators empirically, indicators can also be studied from a theoretical point of view. In theoretically oriented research, indicators are compared with each other based on their mathematical properties. If an indicator has properties that are considered desirable, this provides support for the indicator, and the other way around, if an indicator has properties that are considered undesirable, this may be a reason for rejecting the indicator. In the literature on bibliometric indicators, theoretical approaches are less commonly used than empirical ones. Nevertheless, in recent years, a considerable number of theoretical studies have appeared, both on the *h*-index (e.g., Marchant, 2009a; Quesada, 2010; Woeginger, 2008)[1] and on other types of indicators (e.g., Albarrán, Ortuño, & Ruiz-Castillo, 2011; Bouyssou & Marchant, 2011a, 2011b; Marchant, 2009b; Palacios-Huerta & Volij, 2004; Ravallion & Wagstaff, 2011; Waltman & Van Eck, 2009a, 2010; Waltman, Van Eck, Van Leeuwen, Visser, & Van Raan, 2011). In the present paper, we build on some of this earlier research.

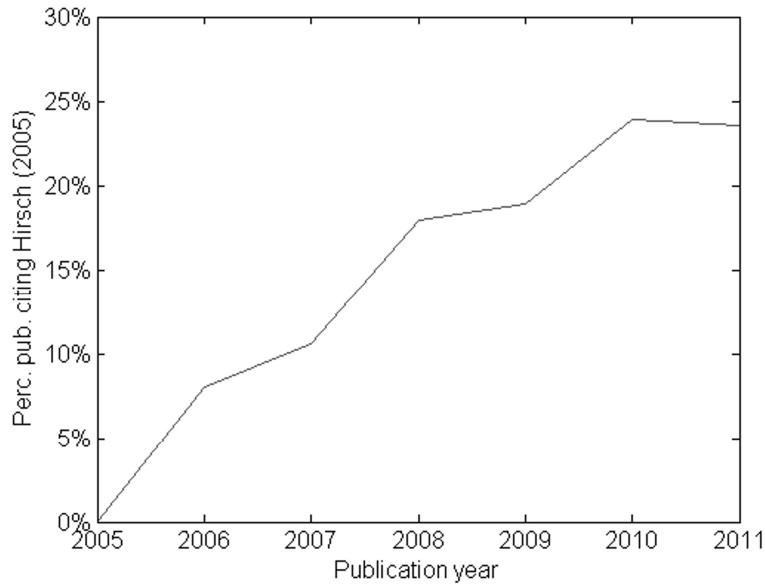

Figure 1. Percentage publications in *Scientometrics* and *Journal of Informetrics* citing Hirsch (2005). (Data retrieved from the Web of Science database on August 6, 2011. Data for 2011 are incomplete.)

The main objective of our paper is to criticize the *h*-index and its variants from a theoretical point of view.[2] Because of the somewhat unsystematic and strongly empirically oriented nature of most *h*-index research, we believe that a fundamental problem of the *h*-index has remained largely unnoticed. We argue that for the purpose of measuring the overall scientific impact of a scientist (or some other unit of analysis) the *h*-index exhibits counterintuitive behavior. More specifically, we assert that the mechanism used by the *h*-index to aggregate publication and citation statistics into a single number leads to an inconsistent way of ranking scientists. The conclusion that we draw from this is that the *h*-index cannot be considered an appropriate indicator of the overall scientific impact of a scientist.

It is not our aim to argue in favor of a single alternative to the *h*-index. Instead, based on recent theoretical insights (Marchant, 2009a, 2009b; Waltman & Van Eck, 2009a), we discuss a large family of bibliometric indicators that do not suffer from the same fundamental problem as the *h*-index. This family of indicators offers a broad range of theoretically well-founded alternatives to the *h*-index. There is one indicator to which we pay special attention. This is the highly cited publications indicator. This indicator has a lot in common with the *h*-index, but unlike the *h*-index it does not produce inconsistent rankings.

---

[1] Another stream of theoretical research, which is less relevant for our present work, studies the properties of the *h*-index in a model-based framework. Examples of this literature include the work of Burrell (2007), Egghe and Rousseau (2006), and Glänzel (2006).

[2] An earlier version of the argument that we are going to present has been published in a short contribution to the *ISSI Newsletter* (Waltman & Van Eck, 2009b).

The organization of this paper is as follows. We first discuss the *h*-index and some of its properties. We then argue that the *h*-index has an inconsistency problem, and we discuss possible alternatives to the *h*-index. We close the paper with some concluding remarks. Because we want the paper to be accessible to a broad audience, we present our theoretical argument completely in intuitive terms. We use almost no formal mathematical notation.

## 2. Definition of the *h*-index

The *h*-index is defined as follows: A scientist has an *h*-index of *h* if *h* of his publications each have at least *h* citations and his remaining publications each have fewer than *h* + 1 citations (Hirsch, 2005). The definition of the *h*-index can also be applied to other units of analysis than scientists, for instance to research groups (Hirsch, 2005; Van Raan, 2006) and journals (Braun, Glänzel, & Schubert, 2006). In this paper, we mostly use scientists as our unit of analysis, but the argumentation that we present applies equally well to other units of analysis.

Figure 2 provides a graphical illustration of the calculation of the *h*-index. The figure shows the distribution of citations over the publications of a scientist. The publications of the scientist are sorted in decreasing order of their number of citations. The figure also shows a 45 degree line through the origin. The *h*-index is obtained by identifying the intersection point of the 45 degree line and the citation curve and by taking the corresponding number of publications. In the example in Figure 2, this yields an *h*-index of six.

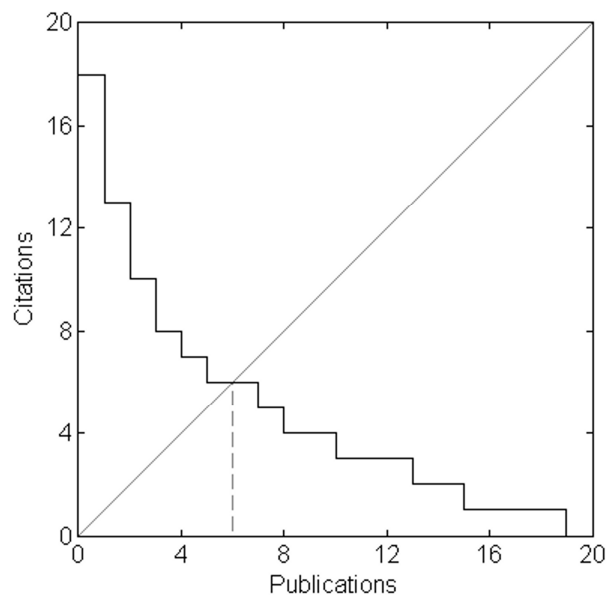

Figure 2. Graphical illustration of the calculation of the *h*-index. The *h*-index is obtained by identifying the intersection point of the 45 degree line and the citation curve and by taking the corresponding number of publications. In this example, the *h*-index equals six.

The *h*-index provides an alternative to simply counting the total number of citations of a scientist. According to Hirsch (2005), a disadvantage of the total number of citations indicator is that it "may be inflated by a small number of 'big hits'" (p. 16569). In other words, the total number of citations indicator is too sensitive to one or a few highly cited publications. The *h*-index does not have this disadvantage. Hirsch' argument in favor of the *h*-index over the total number of citations indicator seems to be accepted by most authors,[3] and it is sometimes added that the *h*-index not only has the advantage of being relatively insensitive to a few highly cited publications but also of being insensitive to large numbers of lowly cited and uncited publications (e.g., Bornmann & Daniel, 2007; Braun et al., 2006).

The *h*-index can also be seen as an alternative to counting the number of highly cited publications of a scientist, where a publication is regarded as highly cited if its number of citations exceeds a certain threshold. According to Hirsch (2005), a disadvantage of the highly cited publications indicator is that the threshold for determining what counts as highly cited "is arbitrary and will randomly favor or disfavor individuals" (p. 16569). In Hirsch' view, the *h*-index has the advantage that it does not depend on a parameter with an arbitrary value. As discussed in an earlier paper (Van Eck & Waltman, 2008), we do not agree with this reasoning. Although the *h*-index does not have any explicit parameters, its definition does involve arbitrariness. For instance, the *h*-index could equally well have been defined as follows: A scientist has an *h*-index of *h* if *h* of his publications each have at least $2h$ citations and his remaining publications each have fewer than $2(h + 1)$ citations. Or the following definition could have been proposed: A scientist has an *h*-index of *h* if *h* of his publications each have at least $h / 2$ citations and his remaining publications each have fewer than $(h + 1) / 2$ citations. A priori, we see no good reason why the original definition of the *h*-index would be better than these two alternative definitions and other similar ones. Because of this, we conclude that, just like the highly cited publications indicator, the *h*-index is subject to arbitrariness. The arbitrariness of the definition of the *h*-index is illustrated graphically in Figure 3. In this figure, like in Figure 2, the *h*-index is obtained by identifying the intersection point of the 45 degree line and the citation curve. Our point is that there is no clear reason for the use of a 45 degree line. As illustrated in Figure 3, one could equally well use, for instance, a 30 degree line or a 60 degree line. Clearly, different lines may cause scientists to be ranked differently. We note that the arbitrariness of the definition of the *h*-index is also recognized by Lehmann, Jackson, and Lautrup (2006, 2008) and Ellison (2010). Lehmann et al. (2008) correctly point out that the *h*-index is based on a comparison of two quantities with different units (i.e., publications and citations). Such a comparison always involves arbitrariness.[4] We will come back to the arbitrariness of the definition of the *h*-index later on in this paper.

The above discussion has focused on a few specific aspects of the *h*-index. The literature on the *h*-index is extensive, and we cannot fully cover it here. For review papers of the *h*-index literature, we refer to Bornmann and Daniel (2007), Alonso, Cabrerizo, Herrera-Viedma, and Herrera (2009), Egghe (2010), and Norris and

---

[3] However, Egghe (2006) considers the insensitivity of the *h*-index to highly cited publications a disadvantage. This has motivated the development of the *g*-index, which seems to be one of the most popular variants of the *h*-index.

[4] Given the arbitrariness of the definition of the *h*-index, there is room for generalizing the index and for proposing simple variants of it. Generalizations of the *h*-index are discussed by Van Eck and Waltman (2008), Deineko and Woeginger (2009), Ellison (2010), and Egghe (2011). Simple variants of the *h*-index are discussed by Kosmulski (2006) and Wu (2010).

Oppenheim (2010). A bibliometric study of the *h*-index literature is reported by Zhang, Thijs, and Glänzel (in press).

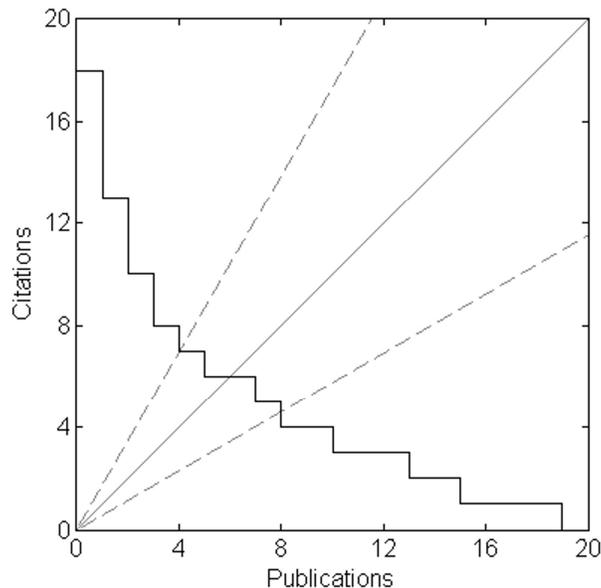

Figure 3. Graphical illustration of the arbitrariness of the definition of the *h*-index. The solid line is a 45 degree line. The dashed lines are a 30 degree line and a 60 degree line. In the calculation of the *h*-index, the 45 degree line is used, but one could equally well use, for instance, the 30 degree line or the 60 degree line.

## 3. Inconsistency of the *h*-index

We now turn to what we consider to be the counterintuitive behavior of the *h*-index. Our aim is to show that the way in which the *h*-index aggregates publication and citation statistics into a single number leads to inconsistent results.

Before going into more detail, it is important to discuss the distinction between size-dependent and size-independent indicators of scientific impact (for a similar distinction, see Waltman & Van Eck, 2009a). Size-dependent indicators never decrease when a scientist obtains an additional publication. Examples of size-dependent indicators are the number of publications, the total number of citations, and the number of highly cited publications of a scientist. The *h*-index is another example. Size-independent indicators, on the other hand, are normalized for the size of someone's oeuvre. For instance, consider a scientist who has three publications, one with *a* citations, one with *b* citations, and one with *c* citations, and consider another scientist who has six publications, two with *a* citations, two with *b* citations, and two with *c* citations. Because the oeuvres of the two scientists differ from each other only in size and not in citation characteristics, a size-independent indicator will have the same value for both scientists. Examples of size-independent indicators are the average number of citations per publication, the median number of citations per publication, and the percentage highly cited publications.

Size-dependent and size-independent indicators serve different purposes. Given a set of publications, a size-dependent indicator is usually interpreted as a measure of the overall scientific impact of the publications, while a size-independent indicator is interpreted as a measure of the average scientific impact per publication. In some cases the use of a size-dependent indicator is more appropriate than the use of a size-independent one, and in other cases it is the other way around. For instance, in the case of the oeuvre of a scientist, measuring the overall impact of the oeuvre is probably more useful than measuring the average impact per publication. This is because measures of the average impact per publication fail to take the productivity of a scientist into account (Hirsch, 2005). In the case of journals, however, things are different. When comparing the impact of journals, one usually does not want size differences to affect the comparison. This means that one needs to use a size-independent indicator (e.g., the impact factor).

Below, we discuss three examples of situations in which the *h*-index is used as a measure of the overall scientific impact of a set of publications. In each example, the *h*-index behaves in a way that we consider counterintuitive. The examples are intended to make clear that, at least for the purpose of measuring the overall impact of a set of publications, the *h*-index provides inconsistent results.

### 3.1. Example 1

In this example, we show that the *h*-index violates the following property:

> *If two scientists achieve the same relative performance improvement, their ranking relative to each other should remain unchanged.*

According to this property, if scientist X is ranked higher than scientist Y and both scientists achieve the same relative performance improvement, then after the performance improvement scientist X should still be ranked higher than scientist Y.

Suppose that scientists X and Y both work at the same university. This university uses the *h*-index as a measure of the overall scientific impact of its scientists. Scientists X and Y have both been active as a researcher for five years. In this five-year period, scientist X has produced twelve publications, nine with twelve citations each and three with four citations each. Scientist Y has produced ten publications, seven with fifteen citations each and three with five citations each. It follows that scientists X and Y have *h*-indices of, respectively, nine and seven (see the left panel of Figure 4). Therefore, based on the *h*-index, the university concludes that scientist X has a larger impact than scientist Y. Suppose next that scientists X and Y keep producing publications at the same rate as before. In terms of citations, their new publications perform equally well as their old ones. This means that after another five-year period scientist X has a total of twenty-four publications, eighteen with twelve citations each and six with four citations each. Scientist Y has a total of twenty publications, fourteen with fifteen citations each and six with five citations each. (For simplicity, we assume that publications from the first five-year period do not receive additional citations.) Hence, the *h*-index of scientist X has increased from nine to twelve, and the *h*-index of scientist Y has increased from seven to fourteen (see the right panel of Figure 4). As a consequence, the university concludes that scientist Y has a larger impact than scientist X. In other words, compared with five years earlier, the ranking of the two scientists has reversed.

In our view, the above result is counterintuitive. Even though in relative terms scientists X and Y have both achieved exactly the same performance improvement (i.e., they have both doubled their publications and citations), their ranking relative to each other has not stayed the same. We consider this very difficult to justify, and we therefore conclude that the *h*-index has ranked scientists X and Y in an inconsistent way.

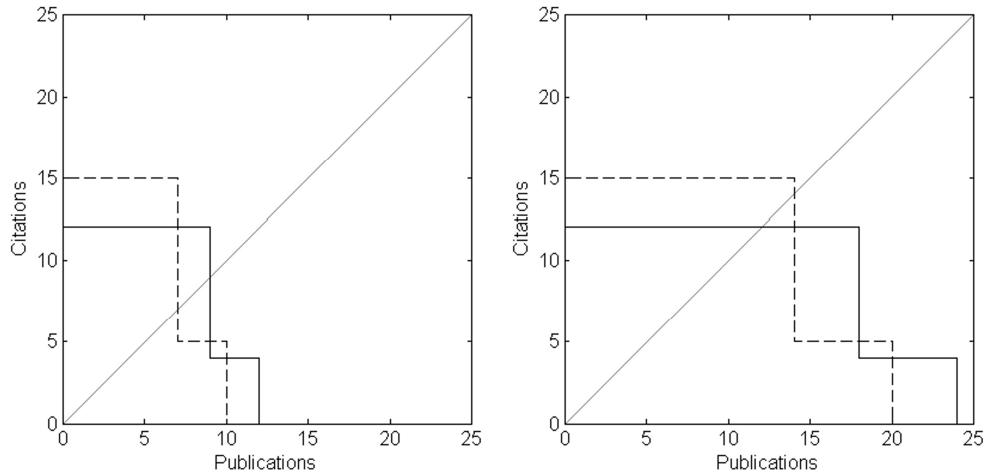

Figure 4. Graphical illustration of the calculation of the $h$-index for scientist X (solid citation curve) and scientist Y (dashed citation curve) in example 1. The left panel shows the situation after the first five-year period. The right panel shows the situation after the second five-year period.

### 3.2. Example 2

We now turn to our second example. In this example, the $h$-index is shown to violate the following property:

*If two scientists achieve the same absolute performance improvement, their ranking relative to each other should remain unchanged.*

This property is quite similar to the property considered in the previous example. The only difference is that the property in the previous example is concerned with relative performance improvements while this property is concerned with absolute performance improvements.

Suppose that scientists X and Y each have seven publications. Scientist X has five publications with five citations each and two publications with two citations each. Scientist Y has four publications with six citations each and three publications with three citations each. The $h$-indices of scientists X and Y then equal, respectively, five and four (see the left panel of Figure 5). Interpreting the $h$-index as a measure of the overall scientific impact of a scientist, it follows that scientist X is ranked higher than scientist Y in terms of overall impact. Suppose next that the two scientists jointly produce two new publications. These publications each receive eight citations. This does not change the $h$-index of scientist X. However, the $h$-index of scientist Y increases from four to six (see the right panel of Figure 5). Hence, after the production of the two joint publications, scientist Y has a higher $h$-index than scientist X. This means that the ranking of the two scientists in terms of overall impact has reversed.

We believe the above result to be counterintuitive. Scientists X and Y have both achieved exactly the same performance improvement (i.e., two publications with six citations each), but despite of this their ranking has reversed. From the point of view of measuring the overall impact of a scientist, we do not see how this can be justified. Like in the previous example, our conclusion therefore is that the $h$-index has provided inconsistent rankings of scientists X and Y.

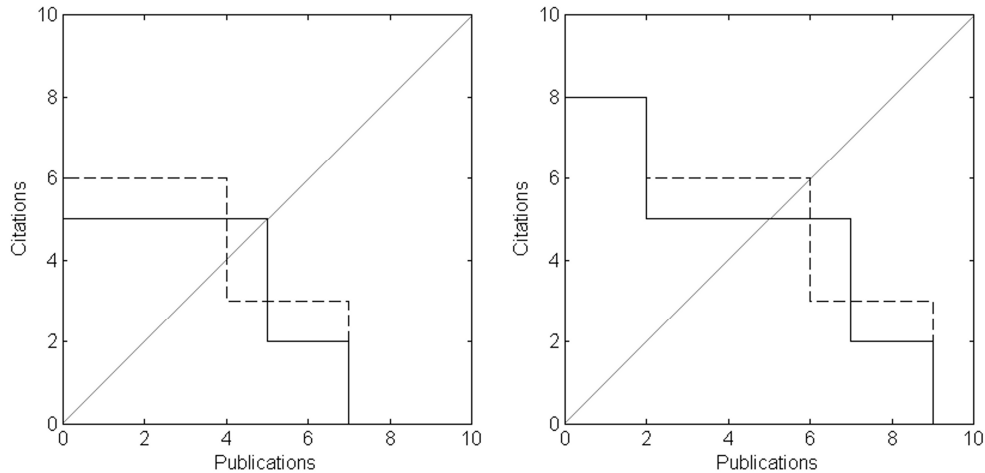

Figure 5. Graphical illustration of the calculation of the *h*-index for scientist X (solid citation curve) and scientist Y (dashed citation curve) in example 2. The left panel shows the initial situation. The right panel shows the situation after the production of two joint publications.

### 3.3. Example 3

The focus of our third example is on consistency between rankings at different levels of aggregation. More specifically, we consider the following property:

*If scientist $X_1$ is ranked higher than scientist $Y_1$ and scientist $X_2$ is ranked higher than scientist $Y_2$, then a research group consisting of scientists $X_1$ and $X_2$ should be ranked higher than a research group consisting of scientists $Y_1$ and $Y_2$.*

We show that the *h*-index does not satisfy this property.

In this example, we use the *h*-index as a measure of the overall scientific impact of both individual scientists and research groups. Suppose we have two research groups, research group X and research group Y, each consisting of two scientists. Research group X consists of scientists $X_1$ and $X_2$, and research group Y consists of scientists $Y_1$ and $Y_2$. Scientists $X_1$ and $X_2$ both have seven publications. Each of their publications has been cited nine times. Scientists $Y_1$ and $Y_2$ both have six publications, and each of their publications has ten citations. The *h*-indices of scientists $X_1$ and $X_2$ then equal seven, while the *h*-indices of scientists $Y_1$ and $Y_2$ equal six (see the left panel of Figure 6). Hence, according to the *h*-index, scientists $X_1$ and $X_2$ have a larger impact than scientists $Y_1$ and $Y_2$. It now seems natural to expect that research group X also has a larger impact than research group Y. However, the curious thing is that according to the *h*-index this is not the case. The *h*-indices of research groups X and Y equal, respectively, nine and ten (see the right panel of Figure 6), which indicates that research group X has a smaller impact than research group Y. This is exactly opposite to what we would expect based on way in which the individual scientists are ranked. We therefore conclude that in our example the *h*-index fails to provide consistent rankings of individual scientists and research groups.

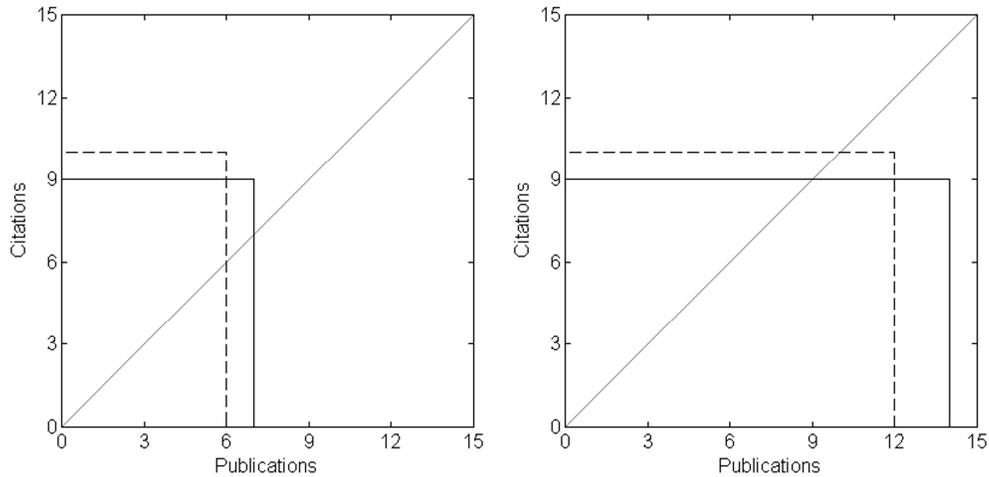

Figure 6. Graphical illustration of the calculation of the $h$-index in example 3. The left panel shows the calculation of the $h$-index for scientists $X_1$ and $X_2$ (solid citation curve) and scientists $Y_1$ and $Y_2$ (dashed citation curve). The right panel shows the calculation of the $h$-index for research group X (solid citation curve) and research group Y (dashed citation curve).

### 3.4. Discussion of the examples

In our view, the above three examples show that, from the perspective of measuring the overall impact of a set of publications, the $h$-index behaves in a counterintuitive way. The mechanism used by the $h$-index to aggregate publication and citation statistics into a single number leads to inconsistent results. Because of this, our conclusion is that the $h$-index cannot be considered an appropriate indicator of the overall scientific impact of a set of publications.

Given the extensive literature on the $h$-index, it is remarkable that the inconsistent nature of the index has remained largely unnoticed. To the best of our knowledge, apart from our own contributions (Waltman & Van Eck, 2009a, 2009b), the only work in which the inconsistency problem of the $h$-index is discussed are papers by Marchant (2009a) and Bouyssou and Marchant (2011b).[5] Marchant (2009a) points out that the $h$-index violates the property that we have discussed in our second example above. He concludes that "the ranking based on the $h$-index is in many circumstances probably not reasonable" (p. 335). Building on our own earlier work (Waltman & Van Eck, 2009a), Bouyssou and Marchant provide a mathematical analysis of the close relationship between the properties discussed in our second and third example above.

It is important to note that the inconsistency problem discussed above affects not only the $h$-index but also all kinds of variants, extensions, and generalizations of this index. For instance, variants of the $h$-index such as the $g$-index (Egghe, 2006), the $h(2)$-index (Kosmulski, 2006), and the $w$-index (Wu, 2010) suffer from similar counterintuitive behavior as the $h$-index itself. The same holds for generalizations of the $h$-index like those proposed by Van Eck and Waltman (2008) and Deineko and Woeginger (2009). Some indicators that are not related to the $h$-index also have an inconsistency problem. Examples include the total number of citations of a scientist's $n$ most highly cited publications and the total number of citations of a scientist's $x$% most highly cited publications. For all these indicators, it is possible to construct examples similar to the above ones that show counterintuitive behavior.

---

[5] Rousseau (2008) provides a short discussion of the paper by Marchant (2009a). The inconsistency problem of the $h$-index is also mentioned very briefly in a book review by Van Raan (2010).

In the literature, various shortcomings of the *h*-index have been discussed. For instance, some authors argue that the *h*-index does not give sufficient credit to highly cited publications (e.g., Egghe, 2006), other authors claim that the index tends to undervalue scientists with a selective publication strategy (e.g., Costas & Bordons, 2007), and still other authors point out limitations of the index in the way in which co-authorship is dealt with (e.g., Egghe, 2008; Hirsch, 2010; Schreiber, 2008). One may wonder whether the above-discussed inconsistency problem is just another shortcoming of the *h*-index or whether in some sense this problem is of a more fundamental nature. In our view, the inconsistency problem is more fundamental. The other problems can all be addressed, at least to a certain degree, by making a suitable modification to the *h*-index, but without leaving the *h*-index framework altogether. The inconsistency problem is more fundamental because it affects not only the *h*-index itself but also all its variants. As will be discussed later on in this paper, solving the inconsistency problem requires a different class of indicators.

### 3.5. Size-independent indicators

What needs to be emphasized is that the above discussion only pertains to size-dependent indicators, that is, indicators aimed at measuring the overall scientific impact of a set of publications. Size-independent indicators, which aim to measure the average scientific impact of the publications in a given set, serve a different purpose and therefore have different requirements.

Consider for instance the average number of citations per publication indicator. Suppose we have two journals, journal X and journal Y. Journal X has five publications, each with six citations. Journal Y has twenty publications, each with five citations. Hence, according to the average number of citations per publication indicator, journal X is ranked higher than journal Y. Suppose next that both journals obtain five new publications. These new publications do not have any citations. Journal X's average number of citations per publication then decreases from six to three, while journal Y's average number of citations per publication decreases from five to four. So the ranking of the two journals reverses. Does this mean that the average number of citations per publication indicator provides inconsistent results? The answer to this question is no. Because the average number of citations per publication indicator aims to measure the average impact of the publications of a journal (rather than the overall impact), the reversal of the ranking of journals X and Y is completely legitimate. Journal Y initially had four times as many publications as journal X. Therefore, journal Y's average impact per publication should be less sensitive to adding new publications than journal X's average impact per publication. In this case, the newly added publications were uncited, which caused both journals' average impact per publication to decrease. However, because of the size difference between the journals, the decrease was more severe in the case of journal X than in the case of journal Y. This caused the ranking of the journals to reverse.

The above example illustrates the difference in the requirements for size-dependent and size-independent indicators. The important thing to keep in mind is that the validity of a size-independent indicator should not be judged based on criteria developed for size-dependent indicators (and the other way around). In the discussion in the next section, we only consider size-dependent indicators. We refer to Bouyssou and Marchant (2011a) and Waltman et al. (2011) for theoretical work on size-independent indicators.

## 4. Alternatives to the *h*-index

Given the inconsistency problem of the *h*-index and many other indicators, one may wonder what kind of alternative indicators can be used that do not have a similar problem. In particular, the question arises whether there are indicators that have similar properties as the *h*-index but that do not suffer from inconsistency problems.

To address these questions, we need to discuss an important recent theoretical paper by Marchant (2009b; see also Waltman & Van Eck, 2009a). Marchant considers a special class of indicators which he refers to as scoring rules.[6] To calculate a scoring rule for a set of publications, one first calculates a score for each individual publication in the set. The score of a publication is determined by the number of citations of the publication. This is done in such a way that an increase in the number of citations of a publication will never lead to a decrease in the score of the publication. After calculating a score for each individual publication, the scoring rule is obtained by calculating the sum of the individual publication scores (or, more generally, by calculating an increasing function of this sum). Somewhat more formally, given a set of $n$ publications with $c_1$, $c_2$, …, $c_n$ citations, a scoring rule is equal to (or, more generally, is an increasing function of)

$$f(c_1) + f(c_2) + \ldots + f(c_n),$$

where $f$ is an increasing function that determines the score of a publication based on the number of times the publication has been cited.

A number of well-known indicators are scoring rules. In particular, the number of publications indicator is a scoring rule in which all publications have the same score, irrespective of their number of citations, and the total number of citations indicator is a scoring rule in which the score of a publication is proportional to the number of citations of the publication. The highly cited publications indicator is a scoring rule as well. In this scoring rule, publications whose number of citations exceeds a certain threshold all have the same positive score and all other publications have a score of zero.

An attractive feature of scoring rules is that they do not suffer from inconsistency problems. Moreover, as shown by Marchant (2009b), almost all indicators that are not a scoring rule do suffer from inconsistency problems.[7] When looking for an alternative to the *h*-index, this essentially means that we can restrict ourselves to scoring rules. In general, indicators that are not a scoring rule will have similar shortcomings as the *h*-index and therefore do not solve the fundamental problem we have with this index.

There are many different scoring rules, and it is not our aim to designate a single scoring rule as the best one. However, we do want to discuss some scoring rules that may serve as suitable alternatives to the *h*-index. As we have discussed, the main advantage of the *h*-index is often claimed to be its relative insensitivity to publications with a very large number of citations, and sometimes also its insensitivity to large numbers of publications with no or almost no citations. A scoring rule that in this respect behaves similarly to the *h*-index is the highly cited publications indicator, that

---

is, the indicator that counts the number of publications with at least a certain number of citations.[8] This indicator only cares about whether a publication counts as highly cited or not. The indicator is insensitive to the exact number of citations of a highly cited publication. Also, like the *h*-index, the highly cited publications indicator is insensitive to large numbers of lowly cited and uncited publications. Another point of similarity between the highly cited publications indicator and the *h*-index is that both indicators have a simple and easy-to-explain calculation. So in many respects the two indicators are similar. However, compared with the *h*-index, the highly cited publications indicator has the advantage that it does not suffer from inconsistency problems. Because of this important advantage, we regard the highly cited publications indicator as a more appropriate indicator of scientific impact than the *h*-index.

The highly cited publications indicator is sometimes argued to have the disadvantage that it depends on an essentially arbitrary parameter for determining which publications count as highly cited and which do not. This is for instance the objection of Hirsch (2005) against the highly cited publications indicator. However, as we have pointed out earlier in this paper, the *h*-index is subject to the same kind of arbitrariness as the highly cited publications indicator. The only difference is that the *h*-index does not have any explicit parameters, which makes it somewhat more difficult to recognize the arbitrary elements in its definition. Because the highly cited publications indicator and the *h*-index are both subject to arbitrariness, we do not consider arbitrariness a good argument for rejecting one indicator in favor of the other.

Apart from the highly cited publications indicator, there are all kinds of other scoring rules that may be used as an alternative to the *h*-index. For instance, one may use a scoring rule in which the score of a publication is given by a (strictly) concave function (such as the square root or the natural logarithm) of the number of citations of the publication (e.g., Lundberg, 2007). Like the *h*-index, such a scoring rule will typically be relatively insensitive to publications with a very large number of citations. Compared with the highly cited publications indicator, a scoring rule that uses a concave function to determine the score of a publication has the advantage that the score of a publication increases in a gradual way as the number of citations of the publication increases. In the case of the highly cited publications indicator, there is an abrupt increase when the number of citations passes the highly cited threshold. This may be considered somewhat unsatisfactory. Clearly, determining the most appropriate score function for a scoring rule is a difficult problem. We refer to the recent work of Ravallion and Wagstaff (2011) for a theoretical framework in which this problem can be explored further.

## 5. Discussion and conclusion

In this paper, we have criticized the *h*-index and its variants from a theoretical point of view. We have argued that for the purpose of measuring the overall scientific impact of a scientist (or some other unit of analysis) the *h*-index behaves in a counterintuitive way. In certain cases, the mechanism used by the *h*-index to aggregate publication and citation statistics into a single number leads to inconsistencies in the way in which scientists are ranked. Our conclusion is that the *h*- index cannot be considered an appropriate indicator of the overall scientific impact of a scientist.

---

[8] For studies on highly cited publications indicators, we refer to Plomp (1990, 1994) and Tijssen, Visser, and Van Leeuwen (2002).

Based on recent theoretical insights (Marchant, 2009a, 2009b; Waltman & Van Eck, 2009a), we have discussed a large family of bibliometric indicators that, unlike the *h*-index, do not suffer from inconsistency problems. The indicators in this family are referred to as scoring rules. It has not been our aim to argue in favor of a single alternative to the *h*-index. However, as we have pointed out, there is one particular scoring rule that has a lot in common with the *h*-index. This is the highly cited publications indicator. Like the *h*-index, the highly cited publications indicator is robust both to publications with a very large number of citations and to publications with no or almost no citations. Yet, compared with the *h*-index, the highly cited publications indicator has the important advantage that it does not suffer from inconsistency problems. Because of this, we believe that the highly cited publications indicator provides an attractive alternative to the *h*-index, at least in situations in which the above-mentioned robustness is regarded as a desirable feature.

It may of course be that the *h*-index has certain good properties which for instance the highly cited publications indicator does not have. We now discuss some suggestions in this direction. Hirsch (2007) argues that compared with other bibliometric indicators the mechanism of the *h*-index is well suited to deal with high impact publications co-authored by scientists with different levels of seniority (or different levels of ability). According to Hirsch, this is because, in a certain sense, senior authors receive more credit from such high impact publications than junior authors (for the full argument, see Hirsch, 2007, p. 19197). Although we consider this an interesting argument, it depends crucially on the assumption that in the case of a high impact publication co-authored by junior and senior scientists most credit should go to the senior authors. Moreover, even if one accepts this assumption, we think it is doubtful whether the somewhat better way of dealing with co-authored publications is worth sacrificing the consistency of one's measurements. A somewhat related argument in favor of the *h*-index is that "the focus of the index shifts in a natural way when comparing researchers at different levels. When comparing young researchers it emphasizes whether they have written a few papers that have had some impact, and when comparing distinguished senior researchers it ignores minor papers and considers only papers that have a substantial number of citations" (Ellison, 2010, p. 2). We agree that this can be seen as an advantage of the *h*-index over for instance the highly cited publications indicator. On the other hand, however, we believe that in the case of young scientists publication and citation statistics are only of limited value, and we therefore do not consider this a very significant advantage of the *h*-index. We also note that the above arguments in favor of the *h*-index pertain specifically to situations in which the *h*-index is used at the level of individual scientists. This means that the arguments cannot serve as a justification for the use of the *h*-index at other levels of aggregation, such as at the level of research groups or journals.

As already mentioned in the beginning of this paper, most of the literature on the *h*-index is empirically oriented. Perhaps the most convincing empirical work in support of the *h*-index is Hirsch' follow-up study on his original *h*-index paper (Hirsch, 2007). In his follow-up study, Hirsch performs a comparison of four bibliometric indicators, namely the number of publications, the total number of citations, the average number of citations per publication, and the *h*-index. Hirsch uses two (relatively small) samples of physicists and looks at publications and citations in two time periods. The focus of Hirsch' study is on the degree to which indicators calculated based on the first time period yield accurate predictions of indicators calculated based on the second time period. Interestingly, the *h*-index not only turns out to be the indicator that is best able to predict its own future value, but it also turns out to be the indicator that is best able to predict the future value of the total number of citations indicator. In our view, Hirsch' study provides a reasonable degree of empirical support to the *h*-index, at least for applications in the field of physics. Unfortunately, apart from the *h*-index, Hirsch' study does not include any other robust indicators, such as the highly cited publications indicator. In future empirical work, we consider it essential that other robust indicators are taken into account as well, especially indicators belonging to the family of scoring rules. This will make it possible to identify indicators that behave in a satisfactory way both from a theoretical and from an empirical point of view.

There is one final remark that we want to make. The idea of the *h*-index is to have a single number that provides a rough approximation of the scientific impact of a scientist. Because of our focus on the *h*-index, we have also adopted this single-indicator viewpoint in this paper. We emphasize, however, that for practical purposes it is usually desirable to have a set of bibliometric indicators, each emphasizing a different aspect of the scientific impact of a scientist. When scientists are being evaluated and compared, it may sometimes be even better to look directly at their citation distributions (e.g., using plots similar to the ones shown in this paper) rather than to focus on indicators derived from these distributions. This always yields the most comprehensive picture of the impact of someone's work as measured by publication and citation data.

## Acknowledgements

We would like to thank Rodrigo Costas, Thierry Marchant, Ronald Rousseau, and Ton van Raan for their feedback on an earlier draft of this paper.